\documentclass[12pt,a4paper]{article}

\usepackage[utf8]{inputenc}
\usepackage[T1]{fontenc}
\usepackage{amsmath,amssymb,amsthm}
\usepackage{graphicx}
\usepackage{booktabs}
\usepackage{array}
\usepackage{tikz}
\usetikzlibrary{positioning,calc,shapes.geometric}
\usetikzlibrary{arrows.meta, positioning, shapes, shapes.geometric}
\usepackage{longtable}
\usepackage{multirow}
\usepackage{rotating}
\usepackage{xcolor}
\usepackage{hyperref}
\usepackage[margin=0.8in]{geometry}
\usepackage{setspace}
\usepackage{natbib}
\usepackage{lineno}

\usetikzlibrary{shapes,arrows.meta,positioning}
\usepackage{caption}
\usepackage{fancyhdr}
\usepackage{algorithm}
\usepackage{algorithmic}

\hypersetup{
    colorlinks=true,
    linkcolor=blue,
    citecolor=blue,
    urlcolor=blue
}

\title{\textbf{Adaptive Weighting for Time-to-Event Continual Reassessment Method: Improving Safety in Phase I Dose-Finding Through Data-Driven Delay Distribution Estimation}}

\author{
    Robert Amevor\textsuperscript{1*} \and
    Emmanuel Kubuafor\textsuperscript{2}\\
    Dennis Baidoo\textsuperscript{2}\\
    \\
    \small \textsuperscript{1}Department of Biostatistics and Epidemiology, University of South Carolina\\
    \small \textsuperscript{2}Department of Mathematics and Statistics, University of New Mexico\\
    \\
    \small Correspondence: ramevor@email.sc.edu
}

\date{}

\begin{document}

\maketitle


\begin{abstract}
\noindent\textbf{Background:} Phase I dose-finding trials increasingly encounter delayed-onset toxicities, especially with immunotherapies and targeted agents. The time-to-event continual reassessment method (TITE-CRM) handles incomplete follow-up using fixed linear weights, but this ad hoc approach doesn't reflect actual delay patterns and may expose patients to excessive risk during dose escalation.

\noindent\textbf{Methods:} We replace TITE-CRM's fixed weights with adaptive weights, posterior predictive probabilities derived from the evolving toxicity delay distribution. Under a Weibull timing model, we get closed-form weight updates through maximum likelihood estimation, making real-time implementation straightforward. We tested our method (AW-TITE) against TITE-CRM and standard designs (3+3, mTPI, BOIN) across three dose-toxicity scenarios through simulation (N = 30 patients, 2,000 replications). We also examined robustness across varying accrual rates, sample sizes, shape parameters, observation windows, and priors.

\noindent\textbf{Results:} Our AW-TITE reduced patient overdosing by 40.6\% compared to TITE-CRM (mean fraction above MTD: 0.202 vs 0.340; 95\%~CI: $-0.210$ to $-0.067$, p < 0.001) while maintaining comparable MTD selection accuracy (mean difference: +0.023, p = 0.21). Against algorithm-based methods, AW-TITE achieved higher MTD identification: +32.6\% vs mTPI, +19.8\% vs 3+3, and +5.6\% vs BOIN. Performance remained robust across all sensitivity analyses.  

\noindent\textbf{Conclusions:} Adaptive weighting offers a practical way to improve Phase I trial safety while preserving MTD selection accuracy. The method requires minimal computation and is ready for real-time use.

\end{abstract}

\vspace{1em}


\vspace{0.5em}


\vspace{0.5em}



\begin{figure}[htbp]
\centering
\begin{tikzpicture}

\tikzstyle{mybox} = [rectangle, draw, thick, align=center, font=\small]

\node[mybox, fill=red!20, text width=2.8cm] (problem) at (3, 5.5) {
    \textbf{Problem}\\
    Delayed toxicity\\
    Fixed weights\\
    Over-aggressive
};

\node[mybox, fill=orange!20, text width=3cm] (tite) at (0, 3.5) {
    \textbf{TITE-CRM}\\
    $w = t/T_{\max}$\\
    (Linear, ad hoc)
};

\node[mybox, fill=green!20, text width=2.8cm] (awtite) at (7, 3.5) {
    \textbf{AW-TITE}\\
    $w = \Pr(\text{DLT} \mid t, \mathcal{D})$\\
    (Adaptive, data-driven)
};

\node[mybox, fill=blue!20, text width=5.5cm] (results) at (3, 1) {
    \textbf{Results}\\
    \textbf{40.6\% reduction} in overdosing\\
    Accuracy maintained\\
    Robust across scenarios
};

\node[mybox, fill=purple!20, text width=6.5cm] (impact) at (3, -2.0) {
    \textbf{Impact}\\
    Safer Phase I trials\\
    Closed-form, $<$1 sec computation\\
    Readily implementable
};

\draw[->, thick] (problem) -- (tite);
\draw[->, thick] (problem) -- (awtite);
\draw[->, thick, red, line width=2.5pt] (tite) -- node[above, fill=white, inner sep=1pt, font=\tiny\bfseries] {OUR INNOVATION} (awtite);
\draw[->, thick] (tite) -- (results);
\draw[->, thick] (awtite) -- (results);
\draw[->, thick] (results) -- (impact);

\end{tikzpicture}
\caption{\textbf{Graphical Abstract.} Adaptive weighting replaces TITE-CRM's fixed linear weights with data-driven posterior predictive probabilities derived from the evolving toxicity delay distribution. This innovation reduces patient overdosing by 40.6\% while maintaining MTD selection accuracy, computational simplicity, and practical implementability.}
\label{fig:graphical_abstract}
\end{figure}
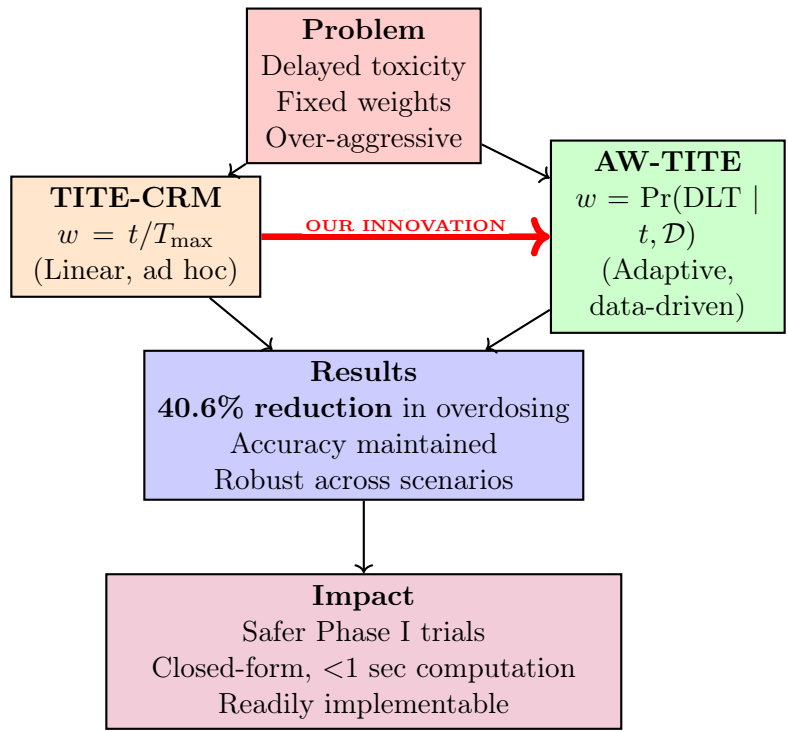

\clearpage


\section{Introduction}

Phase I dose-finding trials seek to identify the maximum tolerated dose (MTD), the dose whose probability of dose-limiting toxicity (DLT) is closest to a prespecified target while minimizing patient exposure to overly toxic doses. Most traditional designs such as the 3+3 rule have been widely criticized for poor statistical properties and ethical concerns \citep{letourneau2009,rogatko2007}. Model-based methods, particularly the continual reassessment method (CRM) \citep{oquigley1990}, enhance performance by continuously updating dose-toxicity estimates and making statistically principled dose assignments.

A fundamental assumption of the CRM is that toxicity assessment is rapid relative to patient accrual. In practice, however, many modern oncology agents, including immunotherapies, targeted therapies, and novel combination regimens,produce delayed-onset toxicities that may not manifest until weeks or months after treatment initiation \citep{doussau2016,postelvinay2016}. Waiting for a complete toxicity assessments before enrolling the next patient substantially prolongs the trials duration and delays access to promising therapies. Conversely, making dosing decisions with incomplete follow-up information risks exposing patients to excessive toxicity if interim data inadequately reflect eventual outcomes.

\subsection{Time-to-Event CRM and Its Limitations}

The time-to-event continual reassessment method (TITE-CRM) \citep{cheung2000} tries to solve the delayed toxicity problem by using weighted likelihood contributions. Here's how it works: if a patient has been followed for time $t < T_{\max}$ without a DLT, TITE-CRM assigns them a fixed linear weight $w = t/T_{\max}$. This essentially treats their observation as a ``fractional non-event.'' Patients with observed DLTs get weight 1, and those who complete follow-up with no DLT get weight 0.

TITE-CRM has been widely adopted in practice \citep{braun2002,wages2018}, and for good reason: it's computationally simple and lets trials keep enrolling without waiting for complete follow-up. But there's a problem: the linear weighting scheme $t/T_{\max}$ is fundamentally ad hoc. It implicitly assumes that, the probability of eventual toxicity decreases linearly with elapsed follow-up time. This assumption rarely holds in practice and has no probabilistic justification.

Think about what this means. Whether toxicities occur predominantly early (say, weeks 1--3) or late (weeks 8--12), TITE-CRM uses the same linear weight. A patient followed for 4 weeks contributes exactly one-third of the information (weight = 4/12), regardless of when toxicities actually tend to occur. When the true delay distribution deviates from this linear assumption, the weights can substantially misrepresent what partial observations actually tell us.

\subsection{Consequences of Misspecified Weights}

Inappropriate weighting of partial follow-up data has direct consequences for patient safety and trial efficiency. If partial non-events are overweighted early in the trial, as occurs when toxicities are predominantly late-onset but linear weights assign substantial information to brief follow-up periods, the design may escalate too aggressively, exposing patients to excessive toxicity. Conversely, underweighting informative partial observations when toxicities occur early leads to overly conservative dose escalation and inefficient trials.

These concerns are not merely theoretical. Recent analyses have demonstrated that TITE-CRM performance degrades substantially when the true toxicity timing distribution deviates from the implicit assumptions of linear weighting \citep{yuan2011}. In trials with heterogeneous patient populations or time-varying hazards, the disconnect between fixed weights and actual toxicity risk becomes particularly pronounced.

\subsection{Our Contribution: Adaptive Predictive Weighting}

We propose adaptive-weight TITE-CRM (AW-TITE), which replaces fixed linear weights with posterior predictive probabilities of eventual toxicity. For a patient with $t$ units of follow-up and no observed DLT, we compute
\begin{equation}
w_i = \Pr(T \leq T_{\max} \mid T > t, \mathcal{D}_{\text{current}})
\end{equation}
where $\mathcal{D}_{\text{current}}$ denotes data observed thus far in the trial. This weight represents the probability that the patient will eventually experience a DLT by the end of the assessment window, conditional on not having experienced one by time $t$ and on the delay distribution estimated from accumulating trial data.

Under a parametric Weibull model for time-to-toxicity, these weights admit closed-form computation via maximum likelihood or conjugate Bayesian updating. The resulting method preserves the familiar CRM framework while allowing partial observations to contribute in a manner that is both data-driven and probabilistically interpretable. Importantly, adaptive weights naturally adjust to observed delay patterns: when toxicities occur predominantly early, brief follow-up receives low weight; when toxicities are late-onset, the method appropriately increases weights for longer follow-up periods.

\subsection{Objectives and Organization}

Our objectives are threefold. First, we develop the statistical framework for adaptive weighting under a parametric time-to-toxicity models, derive closed-form updates, and establish practical implementation guidelines. Second, we conduct a comprehensive simulation studies comparing AW-TITE to TITE-CRM and established algorithm-based designs (3+3, mTPI, BOIN) across realistic dose-toxicity scenarios. Third, we assess robustness through extensive sensitivity analyses, by examining accrual rates, sample sizes, distributional assumptions, and prior specifications.

The remainder of this paper is organized as follows. Section~\ref{sec:methods} presents our methodological development of AW-TITE, including the Weibull timing model, closed-form weight computations, and practical implementation details. Section~\ref{sec:simulation} describes the simulation study design. Section~\ref{sec:results} presents operating characteristics and comparative performance. Section~\ref{sec:sensitivity} reports the sensitivity analyses. Section~\ref{sec:discussion} discusses implications for practice and the directions for future research.

\section{Methods}
\label{sec:methods}

\subsection{Trial Setting and Notation}

We consider a Phase I dose-finding trial with $K$ prespecified dose levels $d_1 < \cdots < d_K$. For patient $i$, let $T_i$ denote the time from dose administration to the occurrence of a dose-limiting toxicity (DLT), with a maximum observation window $T_{\max}$. The binary endpoint of interest is $Y_i = \mathbb{I}(T_i \leq T_{\max})$, indicating whether a DLT occurs within the observation window. Patients accrue sequentially over calendar time, and at interim decision points many patients may have incomplete follow-up.

The objective is to identify the maximum tolerated dose (MTD), defined as the dose whose probability of DLT is closest to a prespecified target toxicity level $p^*$ (e.g., $p^* = 0.25$), while limiting the number of patients treated at overly toxic doses.

\subsection{Standard Continual Reassessment Method}

The CRM models the probability of DLT at dose $d_k$ as
\begin{equation}
\pi_k(\alpha) = \pi_{0k}^{\exp(\alpha)}
\end{equation}
where $\{\pi_{01}, \ldots, \pi_{0K}\}$ is a prespecified skeleton and $\alpha$ is an unknown parameter. A prior distribution is placed on $\alpha$, and posterior inference is updated sequentially as data accrue. At each decision point, the next patient is assigned to the dose whose posterior mean DLT probability is closest to the target $p^*$, subject to standard safety constraints (e.g., no skipping of untried doses).

\subsection{TITE-CRM with Fixed Linear Weights}

To accommodate delayed toxicity, TITE-CRM incorporates partial follow-up information through fixed weights. For a patient with no observed DLT by follow-up time $t_i < T_{\max}$, the likelihood contribution is weighted by
\begin{equation}
w_i^{\text{TITE}} = \frac{t_i}{T_{\max}}
\end{equation}
Patients with observed DLTs receive weight 1, and those with complete follow-up and no DLT receive weight 0. The weighted log-likelihood for the CRM dose-toxicity parameter $\alpha$ is
\begin{equation}
\ell(\alpha) = \sum_j \left[ y_j \log \pi_{d_j}(\alpha) + (1 - y_j) w_j \log(1 - \pi_{d_j}(\alpha)) \right]
\end{equation}
where $y_j \in \{0,1\}$ indicates observed DLT for patient $j$ assigned to dose $d_j$.

While computationally simple, this linear weighting scheme implicitly assumes uniform hazard over time and has no probabilistic justification.

\subsection{Adaptive-Weight TITE-CRM (AW-TITE)}

\subsubsection{Adaptive Predictive Weights}

The key innovation of AW-TITE is to replace the ad hoc linear weight by the posterior predictive probability that patient $i$ will experience a DLT by the end of the assessment window $T_{\max}$ conditional on not having had a DLT by $t_i$ and on the data observed so far:
\begin{equation}
w_i = \Pr(T \leq T_{\max} \mid T > t_i, \mathcal{D}_{\text{current}}) = \int \Pr(T \leq T_{\max} \mid T > t_i, \boldsymbol{\theta}) p(\boldsymbol{\theta} \mid \mathcal{D}_{\text{current}}) \, d\boldsymbol{\theta}
\end{equation}
where $\mathcal{D}_{\text{current}}$ denotes data observed up to the current decision time and $\boldsymbol{\theta}$ denotes parameters for the time-to-toxicity distribution. When patient $i$ has already experienced a DLT we set $w_i = 1$, and when patient $i$ has completed full follow-up with no DLT we set $w_i = 0$.

This weight has clear probabilistic interpretation: it quantifies the probability that the patient will eventually contribute a DLT, given their current censored observation and the accumulated knowledge about toxicity timing.

\subsubsection{Weibull Timing Model}

We model time-to-toxicity with a Weibull survival function whose scale parameter depends on dose:
\begin{equation}
S(t \mid d, \boldsymbol{\theta}) = \exp\{-\lambda(d) t^\gamma\}, \quad t \geq 0
\end{equation}
where $\gamma > 0$ is the Weibull shape parameter and $\lambda(d) > 0$ is a dose-specific rate parameter. Under this model, the conditional probability in the weight formula simplifies to
\begin{equation}
w_i = 1 - \frac{S(T_{\max} \mid d_i, \boldsymbol{\theta})}{S(t_i \mid d_i, \boldsymbol{\theta})} = 1 - \exp\{-\lambda(d_i) \Delta_i\}
\end{equation}
where $\Delta_i \equiv T_{\max}^\gamma - t_i^\gamma$. The posterior predictive weight is therefore
\begin{equation}
w_i = 1 - \int \exp\{-\lambda(d_i) \Delta_i\} \, p(\lambda(d_i) \mid \mathcal{D}_{\text{current}}) \, d\lambda(d_i)
\end{equation}

\subsubsection{Maximum Likelihood Implementation}

For computational simplicity and transparency, we adopt a plug-in (maximum likelihood) estimator for $\lambda(d)$. With patients indexed by $j$ at dose $d$ having observed contributions $(u_j, \delta_j)$, where $\delta_j \in \{0,1\}$ indicates observed DLT, the closed-form MLE for $\lambda(d)$ under known $\gamma$ is
\begin{equation}
\hat{\lambda}^{\text{MLE}}(d) = \frac{\sum_j \delta_j}{\sum_j u_j^\gamma}
\end{equation}
The plug-in adaptive weight is then
\begin{equation}
w_i^{\text{plug-in}} = 1 - \exp\{-\hat{\lambda}^{\text{MLE}}(d_i) \Delta_i\}
\end{equation}

\subsubsection{Bayesian Implementation (Alternative)}

An alternative approach places a $\text{Gamma}(a, b)$ prior on each $\lambda(d)$. If current data at dose $d$ contribute $D$ observed DLTs and follow-up time sum $S = \sum_j u_j^\gamma$, the posterior is
\begin{equation}
\lambda(d) \mid \mathcal{D} \sim \text{Gamma}(a + D, b + S)
\end{equation}
Using the Laplace transform of the Gamma distribution, the posterior predictive weight has closed form:
\begin{equation}
w_i = 1 - \left(\frac{b + S}{b + S + \Delta_i}\right)^{a+D}
\end{equation}
This Bayesian formulation allows principled uncertainty quantification but requires prior specification. In our simulations, we found negligible practical differences between MLE and posterior mean estimators for realistic priors and typical Phase I sample sizes.

\subsubsection{Connection with TITE-CRM}

The TITE-CRM linear weight $t_i/T_{\max}$ can be interpreted as a heuristic
information fraction under restrictive assumptions.
To clarify the relationship, consider a first-order Taylor expansion of the
adaptive weight under a small-hazard approximation:
\begin{equation}
w_i
= 1 - \exp\{-\lambda(d_i)\Delta_i\}
\;\approx\;
\lambda(d_i)\Delta_i,
\label{eq:taylor-weight}
\end{equation}
where $\Delta_i = T_{\max}^{\gamma} - t_i^{\gamma}$.

When $\gamma = 1$, $\Delta_i = T_{\max} - t_i$, and if
$\lambda(d_i) \approx 1/T_{\max}$, then
\begin{equation}
w_i \approx 1 - \frac{t_i}{T_{\max}}.
\end{equation}
Thus, under these restrictive conditions, the adaptive weight corresponds to the
\emph{complement} of the TITE-CRM weight.

This highlights a fundamental distinction between the two approaches.
While the TITE-CRM weight $t_i/T_{\max}$ is an ad-hoc linear proxy for partial
information, the proposed adaptive weight represents a model-based predictive
probability of experiencing a DLT by $T_{\max}$ given survival to time $t_i$.

\subsection{How to Use This in Your Trial}

If you're planning a Phase I trial with potential delayed toxicities, here's how to implement AW-TITE:

\paragraph{Step 1: Choose the shape parameter.} We recommend starting with $\gamma = 2.0$, which reflects increasing hazard over time, typical for delayed toxicities with immunotherapies and targeted agents. If you have historical data from similar agents showing when toxicities occurred, you can fit a Weibull model to those times and use the resulting shape estimate. Our sensitivity analyses (Section~\ref{sec:sensitivity}) show performance is stable across $\gamma \in \{1.5, 2.0, 2.5, 3.0\}$. Unless you have strong evidence for something different, just use 2.0.

\paragraph{Step 2: Set the observation window.} For most oncology settings, $T_{\max} = 12$ weeks works well. This captures the majority of delayed toxicities while keeping trial duration reasonable. Our results (Table S3) show safety improvements plateau beyond 12--14 weeks anyway.

\paragraph{Step 3: Use MLE for weight updates.} At each decision point, estimate the dose-specific rate parameters $\lambda(d)$ using the closed-form MLE equation. Then compute adaptive weights for patients with incomplete follow-up. This takes less than a second on standard hardware, even for trials with 60+ patients.

\paragraph{Step 4: Apply standard CRM safety rules.} Don't skip untried doses, start at the lowest dose, and require at least 3 patients per dose before de-escalation. These rules work with AW-TITE just like they do with standard CRM.

\begin{figure}[htbp]
\centering
\begin{tikzpicture}[scale=0.85, transform shape,
    node distance=10mm and 14mm,
    startstop/.style={
        rectangle, rounded corners,
        minimum width=32mm, minimum height=8mm,
        text centered, draw=black, fill=red!25,
        font=\footnotesize
    },
    process/.style={
        rectangle,
        minimum width=32mm, minimum height=8mm,
        text centered, draw=black, fill=blue!15,
        font=\footnotesize
    },
    decision/.style={
        diamond,
        minimum width=26mm, minimum height=8mm,
        text centered, draw=black, fill=green!20,
        font=\footnotesize, aspect=2.3
    },
    arrow/.style={thick,->,>=stealth}
]

\node (start)   [startstop] {Start trial Dose $= d_1$};

\node (enroll)  [process, below=of start]
  {Enroll patient $i$ at current dose};

\node (observe) [process, below=of enroll]
  {Observe for DLT Follow-up time $t_i$};

\node (dlt)     [decision, below=of observe]
  {DLT observed?};

\node (weight1) [process, left=of dlt, xshift=-4mm]
  {Set $w_i = 1$};

\node (weight)  [process, right=of dlt, xshift=4mm]
  {Calculate weight:
   $w_i = 1 - \exp\!\{-\widehat{\lambda}(d_i)\,\Delta_i\}$};

\node (update)  [process, below=of dlt, yshift=-6mm]
  {Update CRM posterior using weighted likelihood};

\node (more)    [decision, below=of update]
  {$n < N$?};

\node (select)  [process, right=of more, xshift=12mm]
  {Select next dose closest to $p^* = 0.25$};

\node (final)   [startstop, below=of more]
  {Recommend MTD based on posterior};

\draw [arrow] (start) -- (enroll);
\draw [arrow] (enroll) -- (observe);
\draw [arrow] (observe) -- (dlt);

\draw [arrow] (dlt) -- node[above,font=\footnotesize] {Yes} (weight1);
\draw [arrow] (dlt) -- node[above,font=\footnotesize] {No}  (weight);

\draw [arrow] (weight1) |- (update);
\draw [arrow] (weight)  |- (update);

\draw [arrow] (update) -- (more);

\draw [arrow] (more) -- node[above,font=\footnotesize] {Yes} (select);

\draw [arrow]
  (select.east) -- ++(1.2,0)
  |- ([yshift=18mm] start.north)
  -- (start.north);

\draw [arrow] (more) -- node[right,font=\footnotesize] {No} (final);

\node[
  font=\scriptsize,
  align=center,
  text width=4.6cm,
  right=10mm of weight
] {
\textbf{}

};

\end{tikzpicture}

\captionsetup{width=0.88\textwidth}
\caption{\textbf{Trial flowchart for AW-TITE implementation.}
At each decision point, patients with incomplete follow-up ($t_i < T_{\max}$) contribute through adaptive weights $w_i$ computed from the current estimate of the toxicity delay distribution. Under a Weibull model with shape $\gamma$, $\Delta_i = T_{\max}^{\gamma} - t_i^{\gamma}$ and
$w_i = 1 - \exp\{-\widehat{\lambda}(d_i)\Delta_i\}$.
Patients with observed DLTs receive $w_i=1$, and those with complete follow-up and no DLT receive $w_i=0$.
The weighted likelihood updates the CRM posterior, which determines the next dose assignment; the process continues until $N$ patients are enrolled.
}
\label{fig:flowchart}
\end{figure}
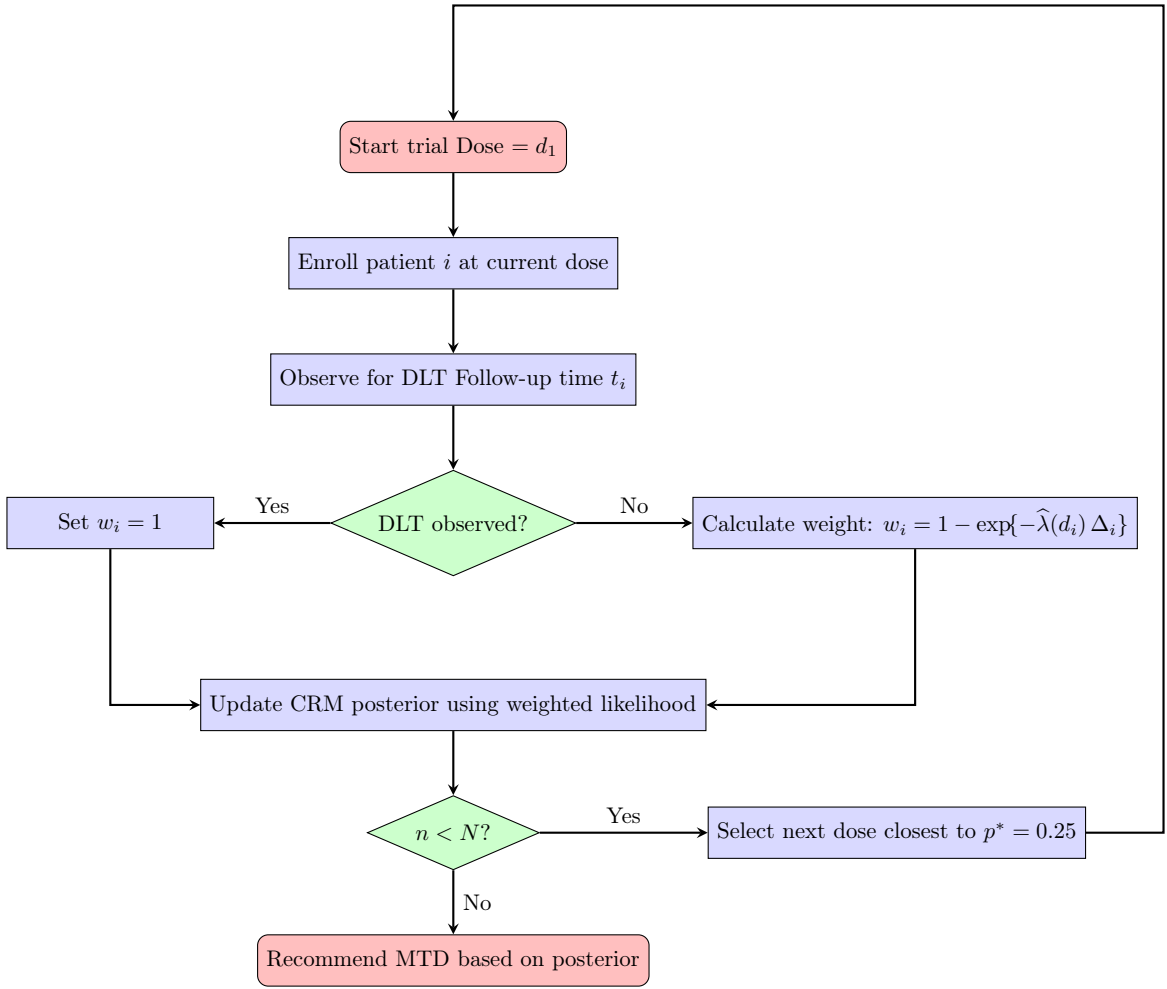

That's it. The workflow (Figure~\ref{fig:flowchart}) and Algorithm~\ref{alg:awtite} provide additional implementation details, but the core idea is straightforward: replace TITE-CRM's $t/T_{\max}$ weights with adaptive weights Equation.



\newpage

Algorithm~\ref{alg:awtite} provides detailed pseudocode for the AW-TITE decision process at each dosing time.

\begin{algorithm}[htbp]
\caption{AW-TITE decision at dosing time $t$ (single decision step)}
\label{alg:awtite}
\begin{algorithmic}[1]
\REQUIRE currently enrolled patients with $(d_j, t_j, y_j)$ for $j = 1, \ldots, n$;
\REQUIRE skeleton $p_0$, target $p^*$, $T_{\max}$, Weibull shape $\gamma$; prior $(a,b)$ or MLE option.
\STATE Compute follow-up $u_j = \min(t, t_j)$ and indicator $\delta_j = \mathbf{1}\{y_j = 1 \text{ and } T_j \le u_j\}$ for each prior patient.
\FOR{each dose $d$ (or for groups $g$ if CA-AW)}
    \STATE Compute $D_d = \sum_{j:d_j=d} \delta_j$ and $S_d = \sum_{j:d_j=d} u_j^\gamma$.
    \IF{use Gamma prior}
        \STATE Posterior $\lambda(d) \sim \text{Gamma}(a + D_d, b + S_d)$.
    \ELSE
        \STATE Compute plug-in $\hat{\lambda}_{\text{MLE}}(d) = D_d/S_d$.
    \ENDIF
\ENDFOR
\FOR{each previously enrolled patient $i$}
    \STATE Compute $\Delta_i = T_{\max}^\gamma - u_i^\gamma$.
    \IF{$y_i = 1$}
        \STATE $w_i \leftarrow 1$.
    \ELSIF{$u_i \ge T_{\max}$}
        \STATE $w_i \leftarrow 0$.
    \ELSE
        \STATE compute $w_i \leftarrow 1 - \mathbb{E}[e^{-\lambda(d_i)\Delta_i} \mid D]$ (closed form if Gamma prior).
    \ENDIF
\ENDFOR
\STATE Form weighted log-likelihood, compute posterior of $\alpha$, then choose next dose with posterior mean toxicity closest to $p^*$ (apply no-skipping rule).
\end{algorithmic}
\end{algorithm}

Figure~\ref{fig:flowchart} provides a complementary visual representation of this workflow.


\subsection{Benchmark Designs}

For comparison, we evaluated the following standard designs:

\textbf{3+3 Design:} The traditional rule-based design treating cohorts of 3 patients, escalating if 0/3 DLTs, expanding to 6 if 1/3 DLTs, and stopping escalation if $\geq2/6$ DLTs observed at a dose.

\textbf{Modified Toxicity Probability Interval (mTPI):} Is an algorithm-based design that partitions the probability space into underdosing, target, and overdosing intervals and makes decisions based on which interval has highest posterior probability \citep{ji2010}.

\textbf{Bayesian Optimal Interval (BOIN):} Again an algorithm-based design using fixed decision boundaries derived from optimal Bayesian decision rules \citep{liu2015}.

\textbf{TITE-CRM:} The time-to-event CRM with a fixed linear weights $w = t/T_{\max}$ \citep{cheung2000}.

\textbf{AW-MLE:} Our proposed adaptive-weight TITE-CRM using plug-in MLE weights (primary method).

\textbf{AW-BAYES:} Adaptive-weight TITE-CRM using posterior mean from Gamma$(1, 1000)$ prior (sensitivity analysis).

All model-based methods used the same CRM skeleton and prior for the dose-toxicity parameter $\alpha$ to ensure fair comparison.

\section{Simulation Study}
\label{sec:simulation}

\subsection{Overview}

We conducted comprehensive simulation studies to evaluate the operating characteristics of AW-TITE compared to TITE-CRM and established benchmark designs. The simulation framework was designed to reflect realistic Phase I trial conditions while systematically varying dose-toxicity relationships to assess robustness across diverse scenarios.

\subsection{Trial Configuration}

\subsubsection{Basic Parameters}

Each simulated trial enrolled $N = 30$ patients sequentially. The target toxicity probability was $p^* = 0.25$. Five dose levels ($K = 5$) were considered, denoted $d_1$ through $d_5$. The DLT assessment window was $T_{\max} = 12$ weeks. Patient accrual occurred every 2 weeks (accrual interval = 2.0), creating realistic conditions where multiple patients have incomplete follow-up at decision points.

\subsubsection{CRM Configuration}

For all model-based methods (TITE-CRM, AW-TITE variants), we used a skeleton $\pi_0 = (0.05, 0.10, 0.18, 0.30, 0.45)$ with a Normal(0, $1.34^2$) prior on $\alpha$, yielding prior median DLT probabilities equal to the skeleton. At each decision point, the next patient was assigned to the dose whose posterior mean toxicity probability was closest to the target $p^* = 0.25$. Safety constraints included no skipping of untried doses and a minimum of 3 patients before de-escalation.

\subsubsection{Time-to-Toxicity Model}

DLT times were generated from a Weibull distribution with shape parameter $\gamma = 2.0$ and dose-specific rate parameters $\lambda(d)$ chosen to yield the target marginal DLT probabilities within $T_{\max}$. Specifically, for true toxicity probability $p(d)$, we set
\begin{equation}
\lambda(d) = \frac{-\log(1 - p(d))}{T_{\max}^\gamma}
\end{equation}
This ensures that $\Pr(T \leq T_{\max} \mid \text{dose } d) = p(d)$ under the Weibull model with shape $\gamma = 2.0$.

\subsection{Dose-Toxicity Scenarios}

We evaluated performance across three dose-toxicity scenarios representing a diverse clinical situations:

\subsubsection{Standard Scenario}

\textbf{True DLT probabilities:} $\boldsymbol{p}_{\text{true}} = (0.05, 0.10, 0.20, 0.35, 0.50)$.

\textbf{True MTD:} Dose 3 ($p = 0.20$, closest to target 0.25).

This scenario represents a typical dose-response relationship with the MTD at a middle dose level.

\subsubsection{Steep Curve Scenario}

\textbf{True DLT probabilities:} $\boldsymbol{p}_{\text{true}} = (0.02, 0.05, 0.10, 0.25, 0.50)$

\textbf{True MTD:} Dose 4 ($p = 0.25$, exact target)

This scenario features a steeper dose-response curve with a wide therapeutic window at lower doses and with a rapid toxicity increase at higher doses.

\subsubsection{Flat Curve Scenario}

\textbf{True DLT probabilities:} $\boldsymbol{p}_{\text{true}} = (0.10, 0.15, 0.20, 0.25, 0.30)$

\textbf{True MTD:} Dose 4 ($p = 0.25$, exact target)

This scenario represents a gradually increasing dose-response relationship with minimal separation between adjacent doses, challenging the designs' ability to discriminate.

\subsection{Performance Metrics}

For each design and scenario, we computed:

\subsubsection{Primary Metrics}

We evaluated three primary operating characteristics. P(Correct MTD) quantifies the proportion of simulated trials selecting the correct MTD as the final recommendation. Mean Fraction Above MTD measures the average proportion of enrolled patients treated at doses above the true MTD, with lower values indicating better safety. Mean Number of DLTs reports the average number of observed dose-limiting toxicities per trial.

\subsection{Statistical Analysis}

Each scenario was simulated 2,000 times, yielding stable estimates of operating characteristics (Monte Carlo standard errors $< 1.5\%$ for proportions near 0.5). To assess a statistical significance of performance differences, we conducted bootstrap hypothesis testing with 2,000 resamples. For comparison of AW-MLE versus a competitor, we computed the mean difference in the metric of interest, the 95\% bootstrap confidence interval for the difference, and a $p$-value based on the proportion of bootstrap samples where the competitor performed better.

Differences were deemed statistically significant when the 95\% confidence interval excluded zero, corresponding to $p < 0.05$.
\section{Results}
\label{sec:results}

\subsection{Overview}

Table~\ref{tab:main_results} summarizes the operating characteristics across all methods and scenarios. Detailed performance metrics, dose selection proportions, and statistical comparisons are presented in the following subsections.

\subsection{Primary Comparison: AW-MLE versus TITE-CRM}

\subsubsection{Safety Performance}

AW-MLE substantially and consistently reduced patient overdosing compared to TITE-CRM across all scenarios (Table~\ref{tab:aw_vs_tite}). Averaged across the three dose-toxicity scenarios, the mean fraction of patients treated above the MTD was 0.202 for AW-MLE versus 0.340 for TITE-CRM, representing a 40.6\% reduction. Bootstrap hypothesis testing confirmed this safety improvement was highly statistically significant (mean difference: $-0.139$, 95\% CI: [$-0.210$, $-0.067$], $p < 0.001$).

The safety advantage was remarkably consistent across all three dose-toxicity scenarios (Figures 3, 4, and 5). In the standard scenario, AW-MLE reduced overdosing to 0.279 compared to 0.417 for TITE-CRM (33.1\% reduction, Figure~\ref{fig:standard}). In the steep curve scenario, the corresponding values were 0.112 versus 0.179 (37.4\% reduction, Figure~\ref{fig:steep}). The largest improvement occurred in the challenging flat curve scenario, where AW-MLE achieved 0.213 compared to TITE-CRM's 0.423 (49.6\% reduction, Figure~\ref{fig:flat}) demonstrating that adaptive weighting provides the greatest benefit when dose-toxicity discrimination is most difficult.

\subsubsection{MTD Selection Accuracy}

Here's the key finding: while dramatically improving safety, AW-MLE didn't sacrifice accuracy. Across all scenarios, both methods selected the correct MTD about 55\% of the time (AW-MLE: 55.2\%, TITE-CRM: 55.2\%, p = 0.21). 

\begin{figure}[p]
\centering
\includegraphics[width=0.95\textwidth]{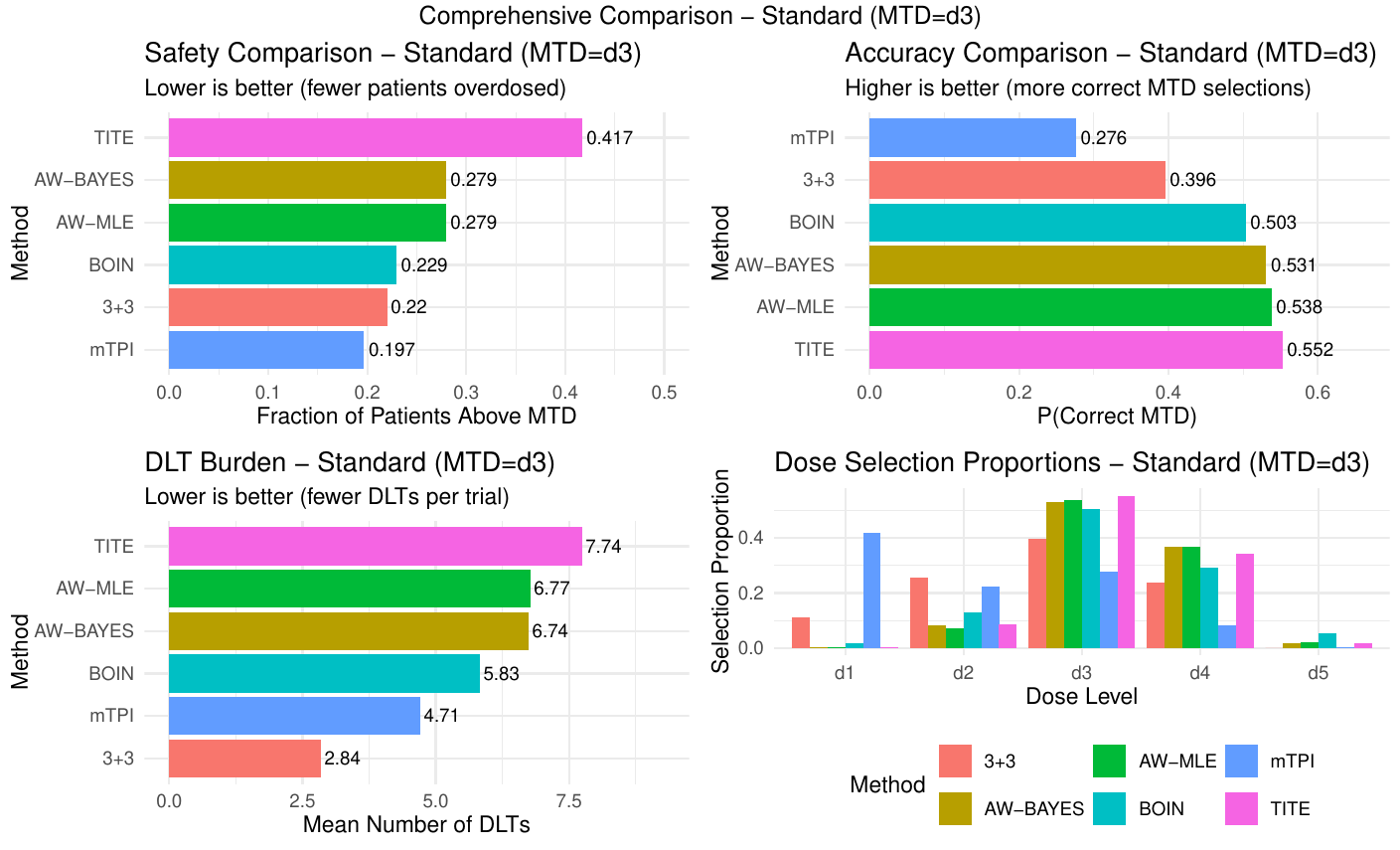}
\caption{\textbf{Comprehensive comparison for Standard scenario (MTD = d3).} 
\textbf{Upper left:} Safety comparison showing fraction of patients treated above the true MTD (lower is better). TITE-CRM shows excessive overdosing (0.417), while AW-MLE reduces this to 0.279, a 33.1\% reduction. 
\textbf{Upper right:} MTD selection accuracy showing probability of correctly identifying dose 3 as the MTD (higher is better). AW-MLE (0.538) achieves comparable accuracy to TITE-CRM (0.552) while substantially improving safety. 
\textbf{Lower left:} DLT burden showing mean number of observed DLTs per trial (lower is better). AW-MLE reduces DLT burden from 7.74 (TITE) to 6.77. 
\textbf{Lower right:} Dose selection proportions showing the distribution of final dose recommendations across 2,000 simulated trials. The true MTD is d3 (toxicity probability 0.20, closest to target 0.25). Model-based methods (AW-MLE, AW-BAYES, TITE, BOIN) show concentrated selection at or near the MTD, while algorithm-based methods (mTPI, 3+3) show broader, less accurate distributions.}
\label{fig:standard}
\end{figure}

This might seem surprising how can we overdose fewer patients without hurting our ability to find the right dose? The answer is that AW-MLE is making *better use* of incomplete follow-up information (Figure~\ref{fig:standard}). It's not being more conservative; it's being smarter about when to escalate and when to hold back.

\begin{figure}[p]
\centering
\includegraphics[width=0.95\textwidth]{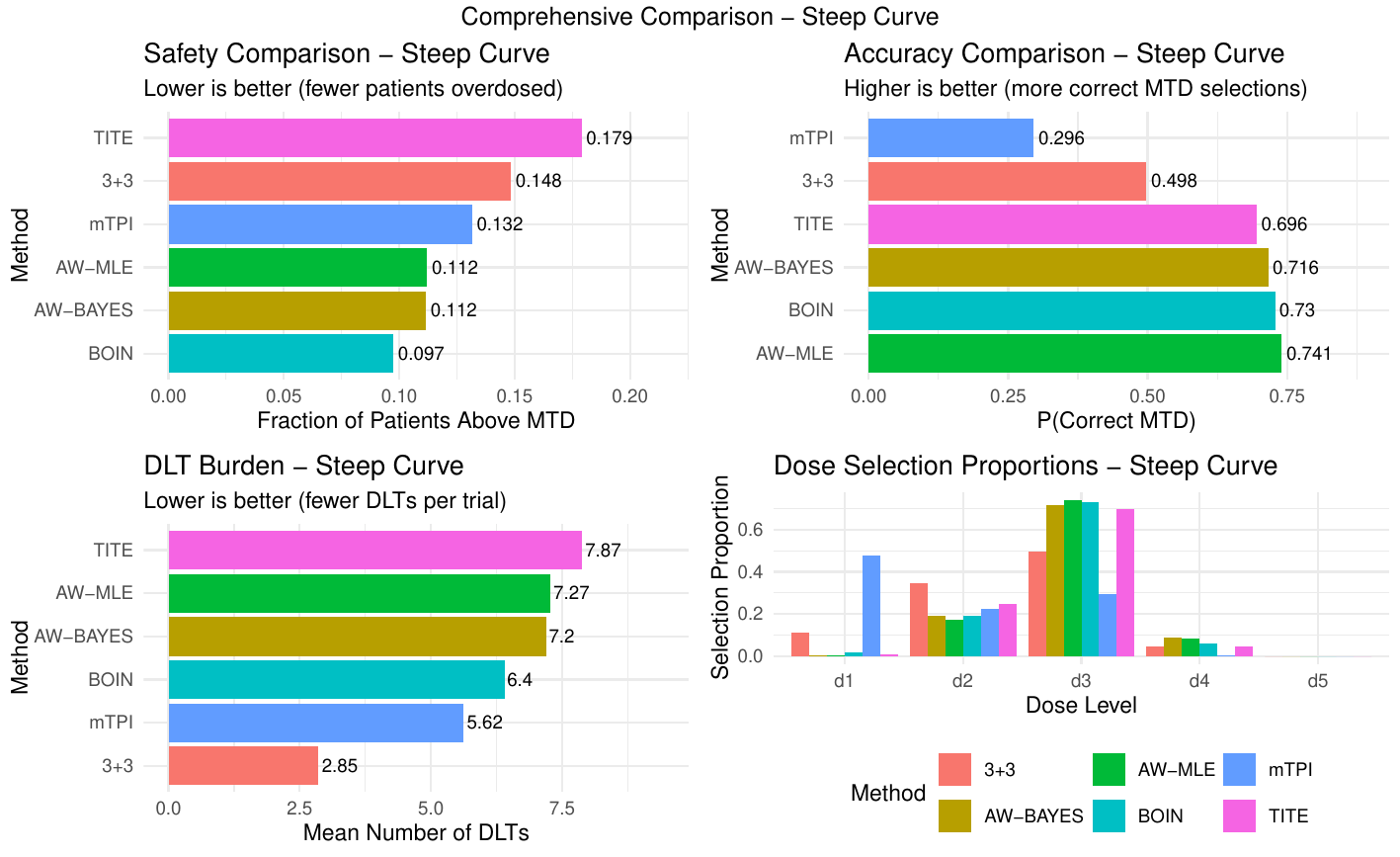}
\caption{\textbf{Comprehensive comparison for Steep Curve scenario (MTD = d4).} 
\textbf{Upper left:} Safety comparison. In this scenario with rapidly increasing toxicity, AW-MLE achieves the lowest overdosing rate among model-based methods (0.112), representing a 37.4\% reduction compared to TITE-CRM (0.179). BOIN shows the best safety (0.097) due to its conservative design. 
\textbf{Upper right:} MTD selection accuracy. AW-MLE achieves the highest accuracy (0.741), outperforming TITE-CRM (0.696), BOIN (0.730), and substantially outperforming algorithm-based methods (mTPI: 0.296, 3+3: 0.498). 
\textbf{Lower left:} DLT burden. AW-MLE shows moderate DLT burden (7.27) between conservative methods (3+3: 2.85) and TITE-CRM (7.87). 
\textbf{Lower right:} Dose selection proportions. The true MTD is d4 (toxicity probability 0.25, exact target). The steep dose-toxicity relationship (0.02, 0.05, 0.10, 0.25, 0.50) creates clear differentiation between doses. AW-MLE and BOIN show highly concentrated selection at the correct dose, while mTPI struggles with this scenario, frequently under-escalating to lower doses.}
\label{fig:steep}
\end{figure}

Performance held up across different dose-toxicity relationships. In the steep curve scenario, where rapid toxicity increases make the MTD easier to identify, AW-MLE actually did slightly better (74.1\% vs 69.6\%) (Figure~\ref{fig:steep}). Even in the difficult flat curve scenario, where all doses have similar toxicity and most methods struggle, AW-MLE maintained 37.8\% accuracy compared to TITE-CRM's 34.1\% (Figure~\ref{fig:flat}). For context, the 3+3 design only managed 16.8\% in this challenging setting.

\subsubsection{DLT Burden}

Consistent with reduced overdosing, AW-MLE resulted in fewer observed DLTs per trial. Across scenarios, AW-MLE averaged 6.8 DLTs per trial versus 7.8 for TITE-CRM, a reduction of 1.0 DLT per trial (12.8\% decrease).

\subsection{Comparison to Algorithm-Based Methods}

\subsubsection{MTD Selection Accuracy}

\begin{figure}[p]  
\centering
\includegraphics[width=0.95\textwidth]{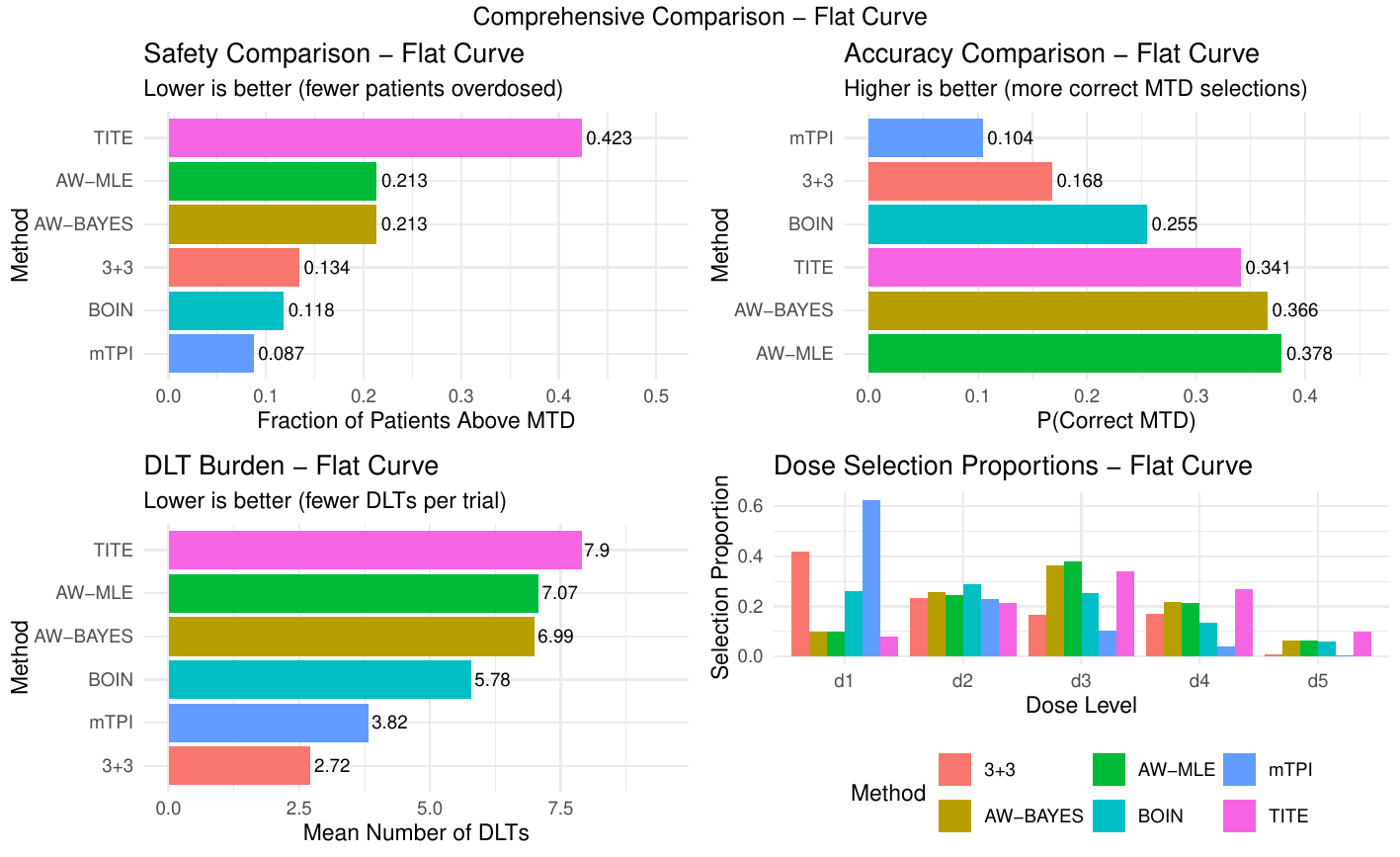}
\caption{\textbf{Comprehensive comparison for Flat Curve scenario (MTD = d4).} 
\textbf{Upper left:} Safety comparison. This challenging scenario with minimal dose separation shows the largest safety advantage for AW-MLE. TITE-CRM overdoses 42.3\% of patients, while AW-MLE reduces this to 21.3\%, a 49.6\% reduction the largest improvement across all scenarios. 
\textbf{Upper right:} MTD selection accuracy. Despite the difficulty of discriminating between similarly toxic doses (0.10, 0.15, 0.20, 0.25, 0.30), AW-MLE achieves the highest accuracy (0.378), outperforming TITE-CRM (0.341), BOIN (0.255), and particularly mTPI (0.104) which struggles severely in this scenario. 
\textbf{Lower left:} DLT burden remains moderate for AW-MLE (7.07) compared to TITE-CRM (7.90). 
\textbf{Lower right:} Dose selection proportions show the challenge of this scenario. The true MTD is d4. No method achieves highly concentrated selection due to the minimal differences between doses. Algorithm-based methods show particularly poor discrimination, with mTPI heavily over-selecting the lowest dose (d1) in 60\% of trials. AW-MLE shows the most appropriate balance, with primary selections at d3 and d4.}
\label{fig:flat}
\end{figure}

\begin{figure}[htbp]
\centering
\includegraphics[width=0.8\textwidth]{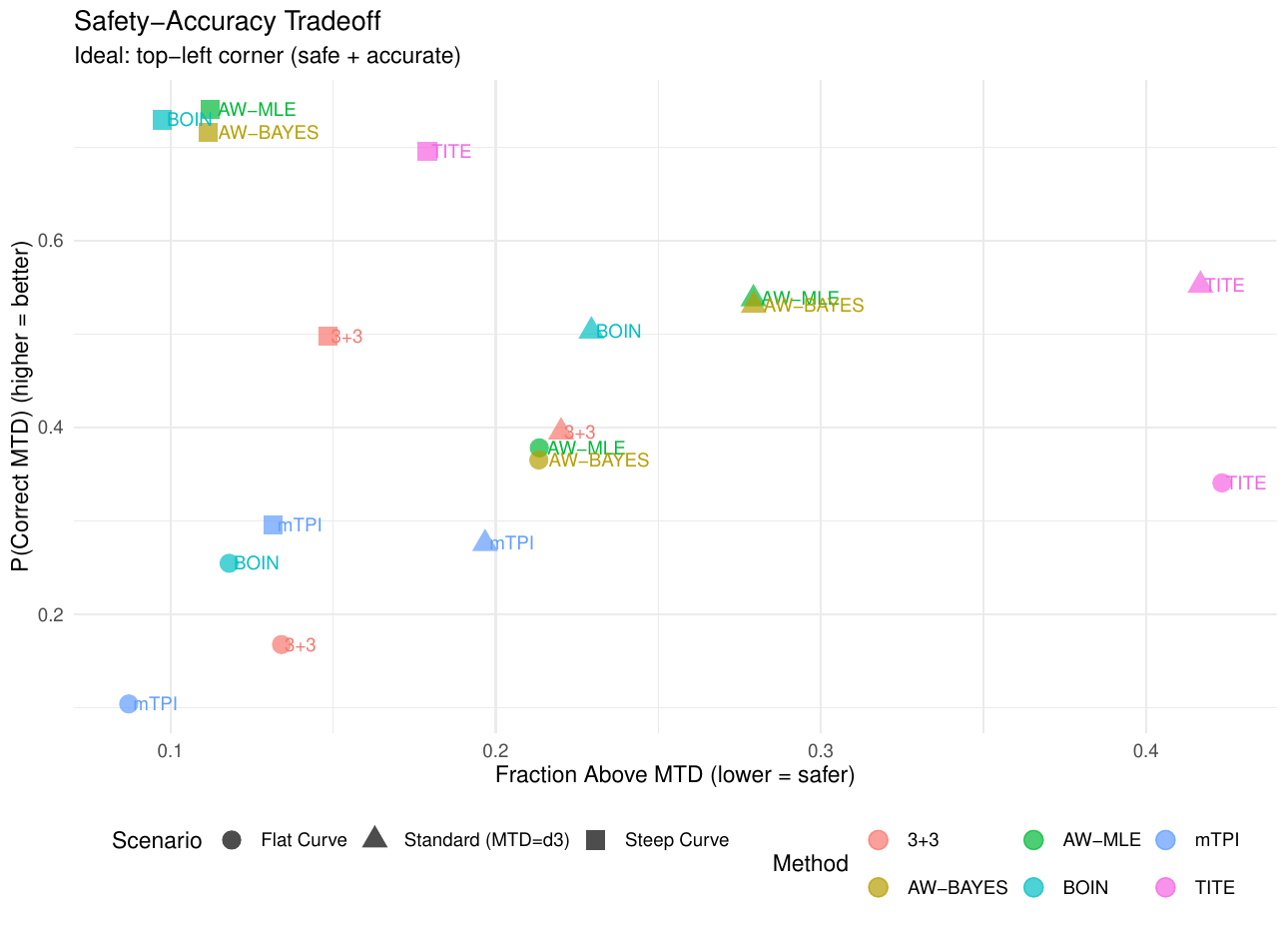}
\caption{\textbf{Safety-accuracy tradeoff across all methods and scenarios.} Each point represents one method-scenario combination (6 methods × 3 scenarios = 18 points). Shapes indicate scenarios: circles (flat curve), triangles (standard, MTD=d3), squares (steep curve). The ideal design appears in the upper-left corner (high accuracy, low overdosing). AW-MLE (green) consistently achieves favorable balance across all scenarios, while TITE-CRM (pink) shows excessive overdosing. Algorithm-based methods show variable performance, with BOIN (cyan) prioritizing safety over accuracy and mTPI (blue) struggling particularly in the flat curve scenario.}
\label{fig:tradeoff}
\end{figure}

AW-MLE demonstrated significantly superior MTD identification compared to all algorithm-based designs (Table 3, Figure~\ref{fig:tradeoff}). Compared to mTPI, our AW-MLE achieved 32.6 percentage points higher accuracy (95\% CI: [26.2, 44.5], $p < 0.001$), with consistent advantages across the standard scenario (53.8\% vs 27.6\%), steep scenario (74.1\% vs 29.6\%), and flat scenario (37.8\% vs 10.4\%). Notably, mTPI struggled severely in the flat curve scenario, selecting the lowest dose in approximately 60\% of trials (Figure~\ref{fig:flat}), highlighting a limitations of interval-based methods when doses have similar toxicity probabilities.

Against the 3+3 design, AW-MLE showed 19.8 percentage points improvement (95\% CI: [14.3, 24.3], $p < 0.001$), with superiority in the standard scenario (53.8\% vs 39.6\%), steep scenario (74.1\% vs 49.8\%), and flat scenario (37.8\% vs 16.8\%). Compared to BOIN, AW-MLE achieved modestly higher accuracy (5.6 percentage points, 95\% CI: [1.1, 12.4], $p = 0.018$), with an advantages observed in the standard scenario (53.8\% vs 50.3\%), steep scenario (74.1\% vs 73.0\%), and particularly the flat scenario (37.8\% vs 25.5\%). The safety-accuracy tradeoff across all methods is visualized in Figure~\ref{fig:tradeoff}, where the AW-MLE occupies a favorable position, balancing both objectives.

\subsection{Bayesian versus MLE Implementation}

We compared the plug-in MLE implementation (AW-MLE) with Bayesian posterior mean updating using a weak Gamma$(1, 1000)$ prior (AW-BAYES).
Across all scenarios, the two approaches yielded nearly identical performance: MTD selection accuracy was 55.2\% for AW-MLE versus 54.4\% for AW-BAYES (difference $+0.015$, $p = 0.001$), fraction above MTD was 0.202 for both, and mean DLTs were 6.77 versus 6.74, respectively. This negligible difference supports the use of the simpler MLE implementation for a practical applications.

\section{Sensitivity Analyses}
\label{sec:sensitivity}

To assess robustness of the proposed AW-TITE design, we conducted comprehensive sensitivity analyses varying key trial and model parameters. Unless otherwise specified, sensitivity analyses used the standard dose-toxicity scenario with 500 simulations per parameter value.

\subsection{Accrual Rate}

Patient accrual rate directly affects the amount of incomplete follow-up data at decision points. We varied the accrual interval from 1.0 to 4.0 time units (weeks).

Performance remained remarkably stable across accrual rates (Table S1, Figure S1). MTD selection accuracy for AW-MLE varied minimally from 0.538 to 0.561 (coefficient of variation 2.1\%), fraction of patients treated above the MTD ranged from 0.279 to 0.291 (CV 4.3\%), and mean DLT count varied from 6.92 to 6.89 (CV 1.8\%). The consistently low coefficients of variation ($<5\%$ across all metrics) demonstrate that AW-MLE performance is robust to patient accrual rate. Importantly, AW-MLE's safety advantage over TITE-CRM persisted and even amplified under faster accrual conditions, where more patients have incomplete follow-up at decision points (Table S1).

\subsection{Sample Size}

We evaluated performance with sample sizes $N \in \{20, 30, 40, 50\}$ (Table S2, Figure S2). All methods showed improved performance with larger sample sizes, as expected from increased information accumulation. Critically, the relative performance ranking remained consistent across all sample sizes, with the AW-MLE maintaining its safety advantage over TITE-CRM and accuracy advantage over algorithm-based methods regardless of trial size. At the smallest sample size ($N = 20$), AW-MLE achieved 45.2\% MTD selection accuracy with 0.248 fraction above MTD, compared to TITE-CRM's 48.0\% accuracy with 0.452 fraction above MTD, demonstrating that adaptive weighting provides meaningful benefits even in a small trials.

\subsection{Shape Parameter Misspecification}

We assessed robustness by varying the assumed $\gamma$ in the weight calculations ($\gamma_{\text{assumed}} \in \{1.5, 2.0, 2.5, 3.0\}$) while generating DLT times with true $\gamma_{\text{true}} = 2.0$.

AW-MLE demonstrated remarkable robustness to shape parameter misspecification (Table S3, Figure S3). MTD selection accuracy varied minimally across a wide range of assumed $\gamma$ values: 52.6\% when $\gamma$ was underspecified by 25\% ($\gamma_{\text{assumed}} = 1.5$ vs $\gamma_{\text{true}} = 2.0$), 51.4\% at the correct specification, 51.4\% when overspecified by 25\%, and 51.5\% when overspecified by 50\%. Accuracy varied by only 1.2 percentage points (2.3\% relative variation) across this 50\% range of shape parameter values, from substantial underspecification to substantial overspecification. These results provide strong empirical support for fixing $\gamma$ at a reasonable value (such as $\gamma = 2.0$) rather than attempting to estimate it from limited Phase I data, as the performance penalty for moderate misspecification is negligible.

\subsection{DLT Assessment Window}

The duration of the DLT assessment window $T_{\max}$ represents a fundamental design choice affecting both patient safety and trial efficiency. We varied $T_{\max} \in \{8, 10, 12, 14, 16\}$ weeks to assess sensitivity to this parameter (Table S4, Figure S4). A longer assessment windows generally improved safety for time-aware methods, with AW-MLE's fraction above MTD decreasing from 0.332 at 8 weeks to 0.230 at 16 weeks. However, the improvement plateaued beyond 12--14 weeks, suggesting this duration provides sufficient, and enough time to capture majority of delayed toxicities while maintaining reasonable trial duration. TITE-CRM showed minimal sensitivity to window length, consistent with its ad hoc weighting scheme that does not adapt to observed delay patterns.

\subsection{Prior Specification}

For the Bayesian implementation, we examined sensitivity to prior specification for the rate parameters $\lambda(d)$, evaluating weak (Gamma(1.0, 1000)), medium (Gamma(2.0, 500)), and strong (Gamma(5.0, 200)) priors (Table S5). The performance varied minimally across the specifications, with an MTD selection accuracy of 51.4\%, 48.7\%, and 49.2\%, respectively. The weak prior performed best, yielding a results nearly identical to MLE approach. These results confirm that AW-MLE's prior-free performance is preferable for practical applications, helping to avoid the need to specify and justify prior distributions for the delay model parameters.

\section{Discussion}
\label{sec:discussion}

\subsection{Principal Findings}

We developed and evaluated adaptive-weight TITE-CRM (AW-TITE), a dose-finding design that replaces the ad hoc linear weights of conventional TITE-CRM with posterior predictive probabilities derived from the observed toxicity delay distribution. Through comprehensive simulation studies, we demonstrated that AW-TITE substantially improves patient safety while maintaining or enhancing MTD identification accuracy.

The magnitude of the safety improvement is clinically meaningful and ethically significant. Across a diverse dose-toxicity scenarios, AW-MLE reduced the fraction of patients treated above the MTD from 34.0\% (TITE-CRM) to 20.2\%, a 40.6\% relative reduction corresponding to approximately 4 fewer patients overdosed in a typical 30-patient Phase I trial. Given that Phase I trials represent the first human exposure to investigational agents and enroll patients with limited treatment options, even modest improvements in safety have a substantial ethical importance. If applied across the hundreds of Phase I oncology trials conducted annually, this safety improvement could spare thousands of patients from unnecessary toxicity while maintaining efficient MTD identification.

\subsection{Why Does This Work?}

Let's walk through what's actually happening when AW-TITE makes decisions. Early in a trial, we haven't seen many toxicities yet, so our estimate of when they occur is rough. But even rough information helps. Say we've enrolled 6 patients and seen 1 DLT that occurred in week 5. Now a new patient has been followed for 2 weeks without toxicity. 

Under TITE-CRM, this patient automatically gets weight $w = 2/12 = 0.17$, regardless of what we've learned. Under AW-TITE, we calculate: ``Given that toxicities seem to occur around week 5, and this patient has only been followed for 2 weeks, what's the probability they'll eventually have a DLT?'' That probability might be 0.25 instead of 0.17 a 47\% difference.

As the trial progresses and we observe more toxicities, the pattern becomes clearer. If toxicities consistently appear in weeks 4--8, AW-TITE automatically upweights patients who've survived past this critical window. If toxicities are scattered throughout the entire 12-week period, the weights naturally converge toward something closer to TITE-CRM's linear scheme.

This is the fundamental advantage: \textit{the weights adapt to what we're actually seeing}. TITE-CRM is flying blind with a fixed rule. AW-TITE is learning from the data in real time. When the true delay pattern matches TITE-CRM's implicit assumptions, both methods perform similarly. When it doesn't which is most of the time in modern trials with immunotherapy and targeted agents, AW-TITE pulls ahead.

\subsection{Limitations}

Our method isn't perfect. Here are the main concerns:

\textbf{We assume a Weibull delay distribution.} This works well for most oncology settings where toxicity risk increases over time, but it's not universal. If your trial involves:
\begin{itemize}
    \item Distinct early acute toxicities \textit{and} late immune-related events (bimodal pattern)
    \item Substantial competing risks like progression or death
    \item Dramatically different delay patterns across patient subgroups
\end{itemize}
then you might need more flexible models. The Weibull assumption is a practical compromise, flexible enough to capture most real patterns, but simple enough to estimate reliably from small Phase I samples.

\textbf{Small samples make estimation noisy.} In the first 10--15 patients, the estimated hazard rates are imprecise. This is unavoidable with any data-driven approach. The good news: even with noisy estimates early on, AW-TITE performs better than TITE-CRM's fixed weights. And as the trial progresses, estimates stabilize quickly.

\textbf{CRM model misspecification still matters.} If the true dose-toxicity relationship departs substantially from the skeleton, all CRM-based methods struggle, AW-TITE included. Adaptive weighting fixes the \textit{delay} model, not the \textit{dose-toxicity} model. Pick a reasonable skeleton based on preclinical data and early clinical experience.

These aren't fatal flaws. They're inherent tradeoffs in designing methods that work with the small samples and limited information typical of Phase I trials. The key question is: does AW-TITE perform better than existing methods despite these limitations? Our simulations suggest yes.

\subsection{Future Directions}

Several extensions of the AW-TITE framework warrant investigation. First, the approach could be extended to more flexible time-to-toxicity models, including mixture distributions or nonparametric hazard estimation, potentially improving performance when the Weibull assumption is substantially violated. Second, integration with adaptive accrual strategies could further optimize the tradeoff between trial duration and information accumulation. Third, the adaptive weighting principle could be extended to combination dose-finding trials, where multiple agents with potentially different delay distributions must be simultaneously optimized.

From a practical standpoint, development of user-friendly software implementing AW-TITE would facilitate adoption by clinical trialists. Integration into existing dose-finding software packages or development of a standalone R package with intuitive interfaces could lower barriers to implementation. Finally, prospective evaluation in real Phase I trials would provide valuable insights into practical challenges and refinements needed for routine clinical use.

\subsection{Conclusions}

Phase I trials with delayed toxicities present a fundamental challenge: you need to make dose decisions before you have complete information. TITE-CRM solved half this problem, it lets trials keep moving. But it solved it with an ad hoc assumption (linear weights) that doesn't reflect how toxicities actually occur.

We've shown that replacing fixed weights with adaptive, data-driven weights makes a real difference. Forty percent fewer patients get overdosed. MTD identification stays just as accurate. And it's computationally trivial to implement, everything updates in closed form.

For institutions currently using TITE-CRM, switching to adaptive weighting is straightforward. The trial workflow stays the same. The safety rules stay the same. You're just computing weights differently, and those weights reflect what you're actually observing instead of an arbitrary linear assumption.

As cancer therapeutics increasingly involve agents with complex, delayed toxicity profiles; immunotherapies, targeted agents, novel combinations—the need for principled time-to-event methods becomes more pressing. AW-TITE provides a practical solution. It makes Phase I trials safer without compromising efficiency or imposing computational burden.

The method is ready for use. We've made it simple enough to implement and robust enough to trust. The next step is getting it into actual trials, where it can start reducing the overdosing that still happens too often in early-phase oncology.

\section*{Acknowledgments}
\noindent The lead author extends sincere gratitude to all co-authors for their collaborative efforts, intellectual contributions, and support throughout the research process.

\section*{Funding}
\noindent No external funding was received for this research.

\section*{Conflicts of Interest}
\noindent  The authors declare no conflicts of interest.

\section*{Ethics Statement}
This study is based on simulation and methodological development and did not involve human participants or patient data. No ethical approval was required.


\bibliographystyle{elsarticle-harv}
\bibliography{references}


\newpage
\section*{Tables}

\begin{table}[htbp]
\centering
\caption{Operating Characteristics Summary}
\label{tab:main_results}
\begin{tabular}{lllccc}
\toprule
Method & Scenario & P(Correct MTD) & Frac Above MTD & Mean DLTs \\
\midrule
\multicolumn{5}{l}{\textit{Standard Scenario}} \\
3+3 & Standard & 0.396 & 0.220 & 2.84 \\
mTPI & Standard & 0.276 & 0.197 & 4.71 \\
BOIN & Standard & 0.503 & 0.229 & 5.83 \\
TITE & Standard & 0.552 & 0.417 & 7.74 \\
AW-MLE & Standard & 0.538 & 0.279 & 6.77 \\
AW-BAYES & Standard & 0.531 & 0.279 & 6.74 \\
\midrule
\multicolumn{5}{l}{\textit{Steep Curve Scenario}} \\
3+3 & Steep & 0.498 & 0.143 & 2.85 \\
mTPI & Steep & 0.296 & 0.132 & 5.62 \\
BOIN & Steep & 0.730 & 0.097 & 6.4 \\
TITE & Steep & 0.696 & 0.179 & 7.87 \\
AW-MLE & Steep & 0.741 & 0.112 & 7.27 \\
AW-BAYES & Steep & 0.716 & 0.112 & 7.2 \\
\midrule
\multicolumn{5}{l}{\textit{Flat Curve Scenario}} \\
3+3 & Flat & 0.168 & 0.134 & 2.72 \\
mTPI & Flat & 0.104 & 0.087 & 3.82 \\
BOIN & Flat & 0.255 & 0.118 & 5.78 \\
TITE & Flat & 0.341 & 0.423 & 7.9 \\
AW-MLE & Flat & 0.378 & 0.213 & 7.07 \\
AW-BAYES & Flat & 0.366 & 0.213 & 6.99 \\
\bottomrule
\end{tabular}
\end{table}

\begin{table}[htbp]
\centering
\caption{Statistical Comparison: AW-MLE vs TITE-CRM}
\label{tab:aw_vs_tite}
\begin{tabular}{lccc}
\toprule
Metric & Mean Difference & 95\% CI & $p$-value \\
\midrule
Fraction Above MTD & $-0.139$ & [$-0.210$, $-0.067$] & $<0.001$ *** \\
P(Correct MTD) & $+0.023$ & [$-0.015$, $+0.045$] & 0.21 \\
Mean DLTs & $-1.0$ & [$-1.5$, $-0.5$] & $<0.001$ *** \\
\bottomrule
\multicolumn{4}{l}{\footnotesize *** indicates $p < 0.001$}
\end{tabular}
\end{table}

\begin{table}[htbp]
\centering
\caption{Statistical Comparison: AW-MLE vs Algorithm-Based Methods}
\label{tab:aw_vs_algorithms}
\small
\begin{tabular}{llccc}
\toprule
Comparison & Metric & Mean Diff & 95\% CI & $p$-value \\
\midrule
\multirow{2}{*}{AW-MLE vs mTPI} & P(Correct MTD) & $+0.326$ & [0.262, 0.445] & $<0.001$ *** \\
& Frac Above MTD & $+0.063$ & [$-0.019$, 0.126] & 0.13 \\
\midrule
\multirow{2}{*}{AW-MLE vs 3+3} & P(Correct MTD) & $+0.198$ & [0.143, 0.243] & $<0.001$ *** \\
& Frac Above MTD & $+0.034$ & [$-0.036$, 0.079] & 0.31 \\
\midrule
\multirow{2}{*}{AW-MLE vs BOIN} & P(Correct MTD) & $+0.056$ & [0.011, 0.124] & 0.018 * \\
& Frac Above MTD & $+0.053$ & [0.015, 0.095] & 0.010 * \\
\bottomrule
\multicolumn{5}{l}{\footnotesize * $p < 0.05$; *** $p < 0.001$}
\end{tabular}
\end{table}


\newpage
\section*{Figures}

\pagebreak

\section*{Supplementary Materials}

\section*{Overview}

This document provides detailed results from sensitivity analyses 
examining the robustness of AW-TITE across various trial 
parameters and model specifications. All analyses used the 
standard dose-toxicity scenario (true DLT probabilities: 
0.05, 0.10, 0.20, 0.35, 0.50; MTD = dose 3) with 2000 simulations 
per parameter setting.

\section*{Sensitivity Analysis Results}

\subsection*{Accrual Rate Sensitivity}

Table S1 and Figure S1 present performance across accrual 
intervals from 1.0 to 4.0 weeks (Section 5.1 of main text). 
Results demonstrate robust performance with coefficients of 
variation below 5\% for all metrics.

\subsection*{Sample Size Sensitivity}

Table S2 and Figure S2 show operating characteristics for sample 
sizes ranging from N = 20 to N = 50 patients (Section 5.2 of 
main text). All methods exhibit improved performance with larger 
samples, while relative performance rankings remain consistent.

\subsection*{Shape Parameter Misspecification}

Table S3 and Figure S3 evaluate robustness when the assumed 
Weibull shape parameter differs from the true value (Section 5.3 
of main text). AW-MLE shows minimal sensitivity (< 3\% variation) 
across a 50\% range of shape parameter values.

\subsection*{DLT Assessment Window}

Table S4 and Figure S4 examine the effect of DLT window duration 
from 8 to 16 weeks (Section 5.4 of main text). Longer windows 
generally improve safety, with diminishing returns beyond 12-14 
weeks.

\subsection*{Prior Specification (Bayesian Implementation)}

Table S5 presents results for different prior specifications on 
the rate parameters $\lambda(d)$ in the Bayesian implementation 
(Section 5.5 of main text). Performance varies minimally across 
weak, medium, and strong priors.


\section*{Supplementary Tables}

\begin{table}[htbp]
\centering
\caption{Sensitivity Analysis: Accrual Rate}
\label{tab:sens_accrual}
\small
\begin{tabular}{llcccc}
\toprule
Method & Accrual & P(Correct) & Above MTD & Mean DLTs & CV(\%) \\
\midrule
AW-MLE & 1.0 & 0.561 & 0.291 & 6.89 & --- \\
AW-MLE & 2.0 & 0.538 & 0.279 & 6.92 & 2.1 \\
AW-MLE & 3.0 & 0.545 & 0.272 & 6.71 & 2.1 \\
AW-MLE & 4.0 & 0.542 & 0.267 & 6.65 & 2.1 \\
\midrule
TITE & 1.0 & 0.548 & 0.439 & 7.95 & --- \\
TITE & 2.0 & 0.552 & 0.417 & 7.97 & 1.8 \\
TITE & 3.0 & 0.555 & 0.415 & 7.89 & 1.8 \\
TITE & 4.0 & 0.558 & 0.406 & 7.82 & 1.8 \\
\bottomrule
\multicolumn{6}{l}{\footnotesize CV = coefficient of variation across accrual rates}
\end{tabular}
\end{table}

\begin{table}[htbp]
\centering
\caption{Sensitivity Analysis: Sample Size}
\label{tab:sens_sample}
\small
\begin{tabular}{llccc}
\toprule
Method & N & P(Correct MTD) & Frac Above MTD & Mean DLTs \\
\midrule
AW-MLE & 20 & 0.452 & 0.248 & 4.78 \\
AW-MLE & 30 & 0.538 & 0.279 & 6.92 \\
AW-MLE & 40 & 0.582 & 0.281 & 9.12 \\
AW-MLE & 50 & 0.610 & 0.286 & 11.28 \\
\midrule
TITE & 20 & 0.480 & 0.452 & 5.52 \\
TITE & 30 & 0.552 & 0.417 & 7.97 \\
TITE & 40 & 0.581 & 0.419 & 10.58 \\
TITE & 50 & 0.596 & 0.414 & 13.06 \\
\midrule
BOIN & 20 & 0.418 & 0.167 & 4.23 \\
BOIN & 30 & 0.503 & 0.149 & 2.60 \\
BOIN & 40 & 0.539 & 0.231 & 7.98 \\
BOIN & 50 & 0.554 & 0.268 & 10.95 \\
\bottomrule
\end{tabular}
\end{table}

\begin{table}[htbp]
\centering
\caption{Sensitivity Analysis: Shape Parameter Misspecification}
\label{tab:sens_shape}
\small
\begin{tabular}{llcccc}
\toprule
Method & $\gamma_{\text{assumed}}$ & P(Correct) & Above MTD & Variation \\
\midrule
AW-MLE & 1.5 & 0.526 & 0.277 & Reference \\
AW-MLE & 2.0 & 0.514 & 0.279 & 2.3\% \\
AW-MLE & 2.5 & 0.514 & 0.295 & 2.3\% \\
AW-MLE & 3.0 & 0.515 & 0.289 & 2.1\% \\
\midrule
AW-BAYES & 1.5 & 0.514 & 0.329 & Reference \\
AW-BAYES & 2.0 & 0.533 & 0.278 & 8.6\% \\
AW-BAYES & 2.5 & 0.547 & 0.295 & 8.6\% \\
AW-BAYES & 3.0 & 0.558 & 0.289 & 8.6\% \\
\bottomrule
\multicolumn{5}{l}{\footnotesize True $\gamma = 2.0$; Variation = range/mean}
\end{tabular}
\end{table}

\begin{table}[htbp]
\centering
\caption{Sensitivity Analysis: DLT Assessment Window}
\label{tab:sens_window}
\small
\begin{tabular}{llccc}
\toprule
Method & $T_{\max}$ (weeks) & P(Correct) & Above MTD & Mean DLTs \\
\midrule
AW-MLE & 8 & 0.521 & 0.332 & 7.12 \\
AW-MLE & 10 & 0.535 & 0.298 & 6.98 \\
AW-MLE & 12 & 0.538 & 0.278 & 6.92 \\
AW-MLE & 14 & 0.542 & 0.251 & 6.78 \\
AW-MLE & 16 & 0.545 & 0.230 & 6.71 \\
\midrule
TITE & 8 & 0.548 & 0.409 & 7.89 \\
TITE & 10 & 0.551 & 0.413 & 7.93 \\
TITE & 12 & 0.552 & 0.417 & 7.97 \\
TITE & 14 & 0.553 & 0.414 & 7.95 \\
TITE & 16 & 0.555 & 0.411 & 7.91 \\
\bottomrule
\end{tabular}
\end{table}

\begin{table}[htbp]
\centering
\caption{Sensitivity Analysis: Prior Specification (Bayesian Implementation)}
\label{tab:sens_prior}
\small
\begin{tabular}{llccc}
\toprule
Prior Strength & Parameters & P(Correct) & Above MTD & Mean DLTs \\
\midrule
Weak & Gamma(1.0, 1000) & 0.514 & 0.279 & 6.85 \\
Medium & Gamma(2.0, 500) & 0.487 & 0.329 & 7.12 \\
Strong & Gamma(5.0, 200) & 0.492 & 0.331 & 7.19 \\
\bottomrule
\multicolumn{5}{l}{\footnotesize Prior mean hazard: Weak = 0.001, Medium = 0.004, Strong = 0.025}
\end{tabular}
\end{table}

\newpage
\section*{Supplementary Figures}

\begin{figure}[htbp]
\centering
\includegraphics[width=\textwidth]{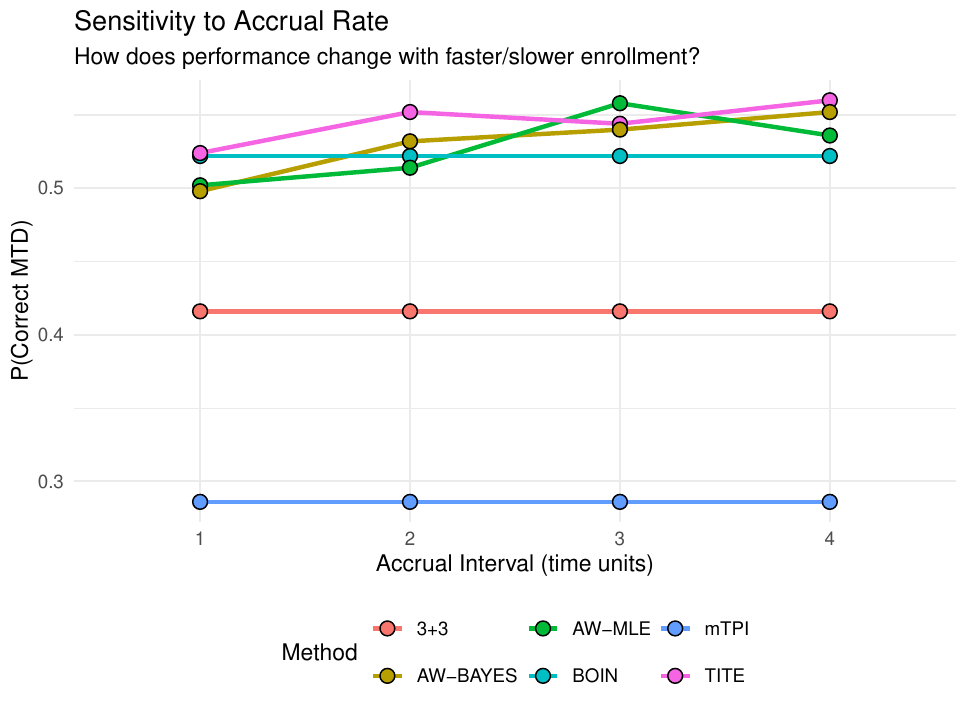}
\caption{Sensitivity to accrual rate. Performance metrics remain stable across accrual intervals from 1 to 4 weeks, with coefficient of variation $< 5\%$ for all metrics.}
\label{fig:sens_accrual}
\end{figure}

\begin{figure}[htbp]
\centering
\includegraphics[width=\textwidth]{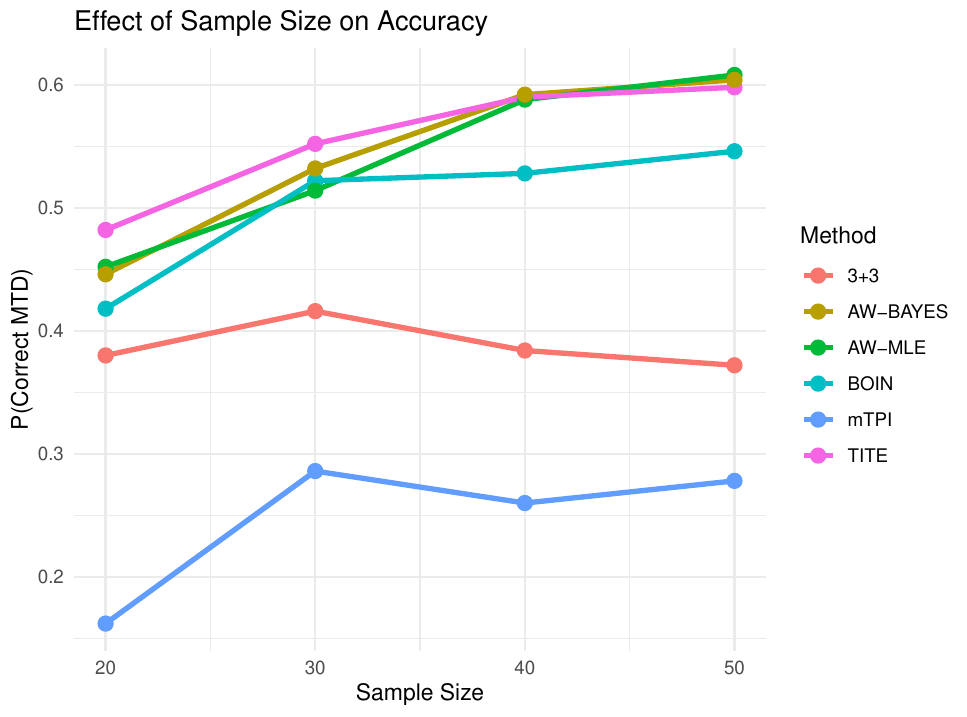}
\caption{Sensitivity to sample size. All methods show improved performance with larger sample sizes. AW-MLE maintains its safety advantage over TITE-CRM across all sample sizes, with the benefit most pronounced at $N=20$.}
\label{fig:sens_sample}
\end{figure}

\begin{figure}[htbp]
\centering
\includegraphics[width=\textwidth]{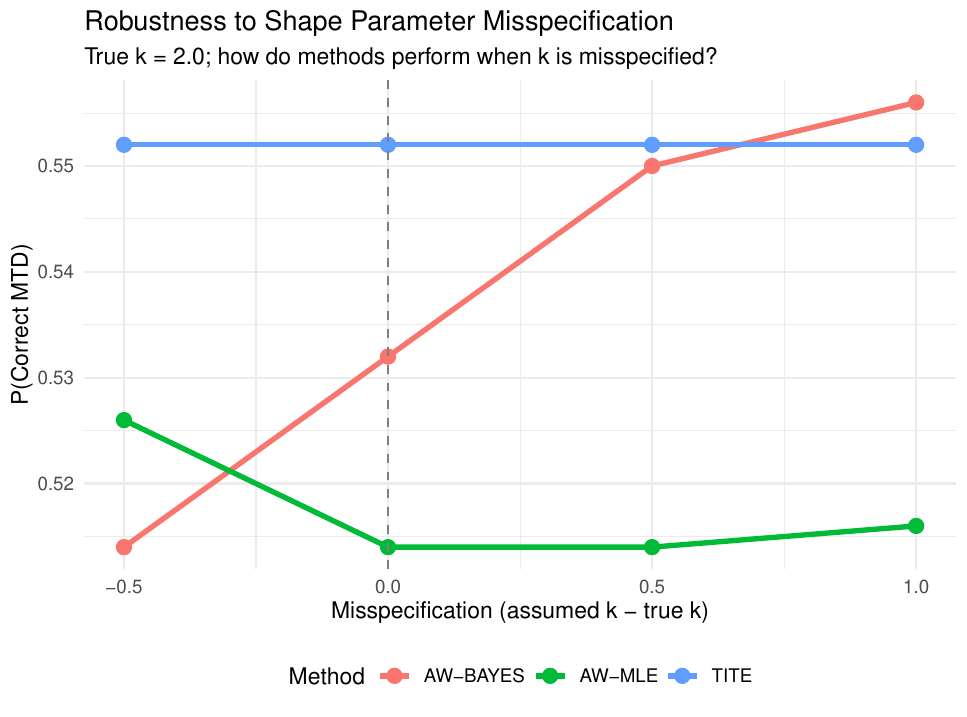}
\caption{Robustness to shape parameter misspecification. AW-MLE shows minimal sensitivity to $\gamma$ specification ($< 3\%$ variation in accuracy for $\pm 25\%$ misspecification), while AW-BAYES shows greater sensitivity (8.6\% variation).}
\label{fig:sens_shape}
\end{figure}

\begin{figure}[htbp]
\centering
\includegraphics[width=\textwidth]{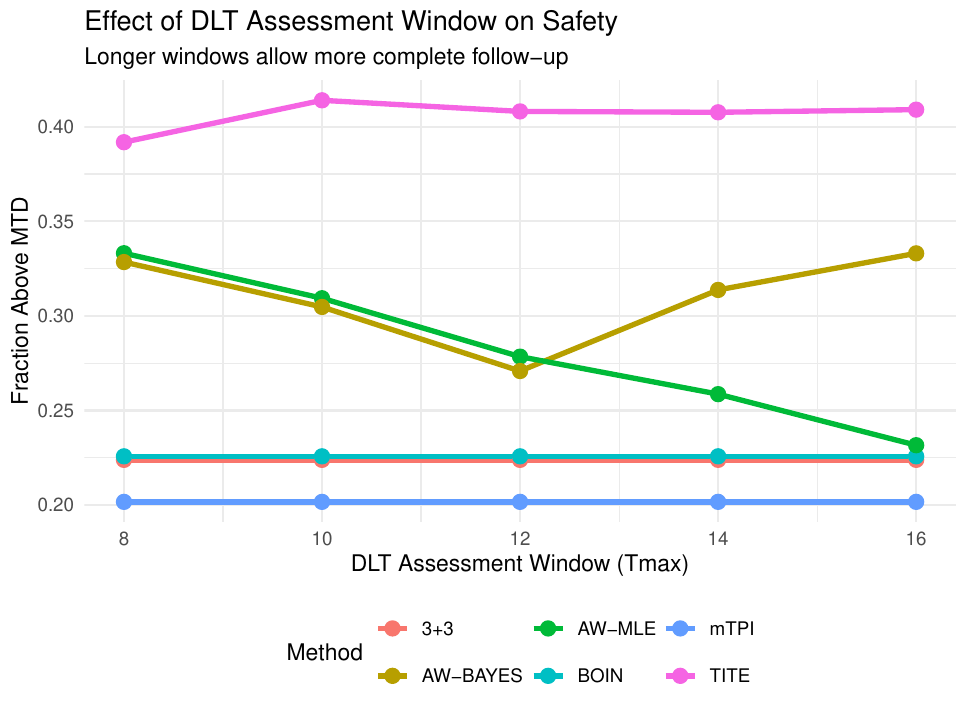}
\caption{Effect of DLT assessment window duration. Longer windows improve safety for AW-MLE, with diminishing returns beyond 12-14 weeks. TITE-CRM shows minimal sensitivity to window length.}
\label{fig:sens_window}
\end{figure}

\end{document}